\documentclass[final,longbibliography,12pt]{revtex4-1}

\usepackage{graphicx}
\usepackage{amssymb}
\usepackage{amsfonts}
\usepackage{amsmath}
\usepackage{slashed}

\newcommand{\be}{\begin{equation}}
\newcommand{\ee}{\end{equation}}
\newcommand{\wf}{wavefunction\;}
\newcommand{\wfs}{wavefunctions\;}
\newcommand{\infl}{influence function\;}
\newcommand{\infls}{influence functions\;}
\newcommand{\rmi}{\mathrm{i}}

\newcommand{\dg}{\dagger}
\newcommand{\fy}{\slashed}

\begin{document}


\title{First Quantized Electrodynamics}

\author{A. F. Bennett}


\email{bennetan@oregonstate.edu}

%

\affiliation{College of Earth, Ocean and Atmospheric Sciences\\
Oregon State University\\104 CEOAS Administration Building\\ Corvallis, OR 97331-5503, USA}

\date{\today}

\begin{abstract}
The parametrized Dirac wave equation represents  position and time as operators, and can be formulated for many particles. It thus provides, unlike field--theoretic Quantum Electrodynamics (QED),  an elementary and unrestricted representation of electrons entangled in space or  time. The parametrized formalism leads directly and without further conjecture to the Bethe-Salpeter equation for bound states. The formalism also yields the Uehling shift of the hydrogenic spectrum, the anomalous magnetic moment of the electron to leading order in the fine structure constant,  the Lamb shift and the axial anomaly of QED. 

\end{abstract}

\maketitle




\section{Introduction}\label{S:int}
Recent observations \cite{Fed11} of photons entangled in time raise the possibility that fermions could be entangled in time.  Quantum Field Theory, as recently shown \cite{Ols12}, does allow a certain sense of timelike entanglement for a massless vacuum state of unspecified spin. However, Quantum Field Theory (QFT) has only limited ability to represent entangled states.  Violations of Bell's inequality in the vacuum state for a massive particle decay exponentially with increasing spacelike separation, the decay scale being the Compton wavelength. If one particle location is subjected to a timelike displacement until the separation is timelike, then violations decay exponentially with that displacement measured again in Compton wavelengths \cite{Summ11}.

The purpose of this analysis is to show that the parametrized Dirac equation proposed by Feynman \cite{Alv98, John69, Fan93} allows the unrestricted space--time entanglement of electrons.  It is also shown that the parametrized formalism leads to the Bethe--Salpeter equation \cite{Bethe51, Salp51,Grei03} for bound states. The formalism further predicts, by use of a partial summation of the Born series, several fundamental phenomena previously predicted \cite{IZ, Wei95, Zee03, Sr07} by field--theoretic or second--quantized Quantum Electrodynamics (QED). Included here are the lowest--order corrections yielding the Uehling potential, the anomalous magnetic moment of the electron, the Lamb shift and the axial anomaly.

The contents of this article are as follows. The parametrized Dirac wave equation is stated in Section  \ref{S:PDWE}. Free wavefunctions, \infls and M\o ller operators for the parametrized Dirac equation are constructed in Section \ref{S:waves}. The constructions are readily extended to several particles, as outlined in Section \ref{S:2part}. Unrestricted entanglement in space--time is shown to be an explicit contingency in the parametrized formalism. The spin--statistics connection has been proved by Jabs \cite{Jabs10} using first quantization, and the line of proof is very briefly described here. Two--particle scattering is defined in terms of M\o ller operators, which lead without further conjecture to the Bethe--Salpeter equation. The construction leading to the M\o ller operators is given in some detail, even though it parallels those found in classic texts  \cite{ BjDr64, Tay72}, since the operators lead  to precise agreement with successful predictions of  field--theoretic QED. Scattering is then simplified to a single particle in an external potential. The first--order scattering of a single particle is outlined in Section \ref{S:scatfirst}, with Mott scattering as an example. The standard one--loop corrections for scattering of arbitrary strength are derived in Section \ref{S:oneloop} by use of a substitution that is accurate for beams undergoing weak scattering. The substitution resembles the QED relation between propagators and vacuum--to--vacuum expectations. The standard QED axial anomaly is derived in Section \ref{S:axanom} for classical fields rather than fields of operators.  The summary in Section \ref{S:summdisc} includes a discussion of the wide utility  of semiclassical theory and the further possibilities for parametrized formalisms.

\section{The parametrized Dirac  wave equation}\label{S:PDWE}
The parametrized Dirac wave equation is stated,  as are its Lagrangian and its discrete symmetries. 
 
\subsection{covariant formulation}\label{S:covform}

For a single spin--1/2 particle  the \wf is a four--spinor  $\psi(x,\tau)$. The event $x$ is in ${\mathbb R}^4$, while the parameter  $\tau$ is an independent variable in ${\mathbb R}$.  The event $x$ is also denoted by $x^\mu$ having indices $\mu=0,1,2,3$, with $x^0=ct$ where $c$ is the speed of light and $t$ is coordinate time. The Lorentz metric $g^{\mu\nu}$ on  ${\mathbb R}^4$ has  signature $(- + + +)$. The position $\bf{x}$  is denoted by $x^j$ having indices $j=1,2,3$. Thus $x=(ct,\bf{x})$.

The parametrized Dirac wave equation for $\psi$ is  
\be\label{PDIR}
\frac{\hbar}{\rmi c}\frac{\partial}{\partial \tau}\psi+\gamma^\mu\Big(\frac{\hbar}{\rmi}\frac{\partial}{\partial x^\mu}-\frac
{e}{c}A_\mu\Big)\psi=0
\ee
where $e$ is the charge of the particle, $c$ is the speed of light and $\hbar$ is the reduced Planck's constant. The $\gamma^\mu$ are the four Dirac matrices,  while the  Maxwell electromagnetic potential $A^\mu(x)$ is  independent of the parameter $\tau$.  The covariance of the theory with respect to the homogeneous Lorentz transformation $(x^\mu)' =\Lambda^{\mu}_{\;\;\nu}x^\nu$ and $ \psi'(x,\tau)=S(\Lambda)\psi(\Lambda^{-1}x,\tau)$ follows for $S(\Lambda)$  generated in the standard way \cite{BjDr64}. No mass constant appears  in (\ref{PDIR}), but masses are introduced through boundary conditions as $\tau\to \pm \infty$. Feynman's development of QED using (\ref{PDIR}) has been reviewed by Garcia Alvarez  and Gaioli \cite{Alv98}. A simple consequence of (\ref{PDIR}) is the identity
\be\label{VCUR}
\frac{\partial}{\partial \tau}\overline{\psi}\psi+\frac{\partial}{\partial x^\mu}j^\mu=0
\ee
where $\overline{\psi} \equiv \psi^\dg\gamma^0$, and the $\tau$--dependent current is $j^\mu= c\overline{\psi}\gamma^\mu \psi$. The indefiniteness of the invariant bilinear form $\overline{\psi}\psi$ has impeded \cite{Barut85, Evans98} the development of the parametrized Dirac formalism as a relativistic extension of quantum mechanics. 

The energy--momentum operator is denoted $p_\mu=(\hbar/\rmi)\partial/\partial x^\mu$, and $\pi_\mu=p_\mu-(e/c)A_\mu$\,. The commutator of the operators $x^\mu$ and $p^\nu$ is 
\be\label{commxp}
[x^\mu,p^\nu] = x^\mu p^\nu -p^\nu x^\mu=\rmi \hbar g^{\mu \nu} \,.
\ee
Henceforth the units are chosen such that  $c=\hbar=1$. The Maxwell field strength tensor is thus $F^{\mu\nu}=\rmi(p^\mu A^\nu-p^\nu A^\mu)$\,. The summation convention is assumed with respect to repetitions of Greek indices such as $\mu =0,1,2,3$\,. The covariant and contravariant indices $\mu,\nu,\dots$ will be omitted wherever convenient, as in $x= x^\mu$\,, $p = p^\mu$\,, $p\cdot x = p^\mu x_\mu$ and $F \cdot F= F_{\mu\nu}F^{\mu\nu}$\,.

The Newton--Wigner position operator \cite{NeWi49} can only localize the position ${\bf x}$ of, say, an electron within a distance of ${\mathcal O} (m_e^{-1})$ where $m_e$ is the conventional rest mass. The operator  is extendable in the parametrized formalism to an event operator $x^\mu$ with a footprint of the same order. However, the formalism does not restrict the state to a single or `sharp' rest mass. Consider a  \wf $\psi(x,\tau) = \exp(\rmi m \tau)\lambda(x)$. The mass $m$ can have infinite range and so perfect localization of $x^\mu$ and of $p^\mu$ is attainable \cite{Ho92,ArHo85}.

\subsection{Lagrangian and discrete symmetries}\label{S:Lagdisc}
The single--particle Lagrangian is
\be\label{LAG}
\mathcal{L}=  \frac{1}{4}\int d^4x\,F \cdot F  + \int d\tau \int d^4x \, \overline{ \psi} \Big(\frac{1}{\rmi}\frac{\partial}{\partial \tau}\psi+\fy{\pi}\psi\Big)\,,   
\ee
where  $d^4x$ is the Lorentz--invariant measure on ${\mathbb R}^4$\,.  The Feynman slash notation is for example $\fy{p}=\gamma^\mu p_\mu$\,, and in particular $\fy{p}\fy{p}=- p \cdot p$\,. Variation of $\mathcal{L}$ with respect to $\overline{\psi}$ yields (\ref{PDIR}), while variation of $\mathcal{L}$ with respect to $A^\mu$ in the Lorenz gauge, where $\partial \cdot A=0$\,, 
yields 
\be\label{MAX}
\square A^\mu= p\cdot p \, A^\mu= -\partial \cdot \partial \,  A =e J^\mu
\ee
which is Maxwell's equation for $A^\mu$. 
The `concatenated' current 
\be\label{concur}
J^\mu(x)=\int  j^\mu(x,\tau)\,d\tau
\ee
in (\ref{MAX}) is divergenceless, satisfying $\partial \cdot J =0$\,.  
 

The discrete symmetries of charge conjugation ${\mathcal C}$, parity ${\mathcal P}$ and time reversal ${\mathcal T}$ act on the single--particle \wfs as follows.
\be 
({\mathcal C}\psi)(x,\tau) = \rmi\gamma^2\psi^*(x,-\tau)          \label{E:disC} 
\ee
\be
({\mathcal P}\psi)(t,{\bf x},\tau) = \gamma^0\psi(t,-{\bf x},\tau) \label{E:disP} 
\ee
\be
({\mathcal T}\psi)(t,{\bf x},\tau)= \rmi \gamma^1 \gamma^3 \psi^*(-t,{\bf x},-\tau) \label{E:disT} 
\ee
\be
(\mathcal {TPC}\psi)(x,\tau) = -
\rmi \gamma^5 \psi(-x,\tau)  \label{E:disCPT}
\ee


In particular, the charge--conjugate \wf $({\mathcal C}\psi)(x,\tau)$ satisfies (\ref{PDIR}), subject to the charge $e$ being replaced with $-e$.

\section{Wave functions and M\o ller operators}\label{S:waves}
The essential constituents of scattering theory are developed in detail for one particle. The details are routine, being close analogs of those in standard quantum mechanics. The major difference is the existence of states having `negative mass' but they do not participate, even virtually, in the scattering process. 

\subsection{free wavefunctions}\label{S:free}

The free \wfs are the solutions of (\ref{PDIR}) in the absence of the electromagnetic potential $A^\mu
$. They have the form
\be\label{FWD}
f^{(+)}_p(x,\tau)=\frac{u(p)}{(2\pi)^2}\exp[\rmi( p \cdot x+ \varphi_p m_p\tau)]
\ee
and
\be\label{BKD}
f^{(-)}_p(x,\tau)=\frac{v(p)}{(2\pi)^2}\exp[\rmi(p \cdot x - \varphi_pm_p\tau)]
\ee
where $p^\mu$ is a constant energy--momentum vector, while $\varphi_p=\theta(p^0)-\theta(-p^0)$  is the sign of $p^0$. That is, $\varphi_p = p^0/E_p$ where  $E_p=|p^0|$. The mass $m_p$, also denoted $m(p)$\,, is the positive square root of $-p\cdot p$.  It is assumed that the states are not superluminal, and so $p\cdot p \le 0$.    The phases of the complex exponentials are $p\cdot x \pm \varphi_p m_p\tau$, hence
\be\label{FB}
\frac{dt}{d\tau}=\pm m_p/E_p
\ee
at constant phase and position, regardless of the value of $\varphi_p$. Recall that the metric signature is $(-+++)$. Thus $f^{(+)}_p$ propagates forward in coordinate time as $\tau$ increases, for both positive--energy ($\varphi_p=+1$) and negative--energy states ($\varphi_p =-1$), while $f^{(-)}_p$ propagates  backward for both kinds of states. 

There are two linearly independent solutions for the four--spinor $u(p)$ and two for the four--spinor $v(p)$. In the rest frame where $p=(p^0,{\bf 0})$ they are
\be\label{LINREST}
u^{(1)}=\left(\begin{matrix}1\\0\\0\\0\end{matrix}\right),\;
u^{(2)}=\left(\begin{matrix}0\\1\\0\\0\end{matrix}\right),\;
v^{(1)}=\left(\begin{matrix}0\\0\\1\\0\end{matrix}\right),\;
v^{(2)}=\left(\begin{matrix}0\\0\\0\\1\end{matrix}\right).
\ee
In an arbitrary frame the solutions for $u$ and $v$ have the $ 4 \times 2$ block forms

\be\label{LIN}
\begin{aligned}
{\bf u}(p)&=\big(u^{(1)}(p),u^{(2)}(p)\big)=K\left(\begin{matrix}(m_p+E_p)I_2\\ \varphi_p{\bf p \cdot \sigma}\end{matrix}\right)\\{\bf v}(p)&=\big(v^{(1)}(p),v^{(2)}(p)\big)=K\left(\begin{matrix}\varphi_p
 {\bf p \cdot  \sigma} \\(m_p+E_p)I_2\end{matrix}\right)
 \end{aligned}
\ee
where $K=[2m_p(m_p+E_p)]^{-1/2}$, and the $\sigma^j$ are the Pauli matrices $(j=1,2,3)$. The symbol $I_n$ denotes the  $n \times n$ unit matrix. The blocks ${\bf u}$ and ${\bf v}
$ are eigenspinors of the `forward' and `backward' projections
\be\label{PROJ}
\Lambda_u(p) \equiv \frac{m_p I_4 - \varphi_p \fy{p}}{2m_p}= {\bf u}\overline{\bf u}\;,\qquad
\Lambda_v(p) \equiv \frac{m_p I_4+  \varphi_p
 \fy{p}}{2m_p}=- {\bf v}\overline{\bf v}
\ee
respectively, and they obey the orthonormality conditions
\be\label{ORTH}
\overline{\bf u}{\bf u}=I_2\,, \qquad \overline{\bf v}{\bf v}=-I_2
\,, \qquad \overline{\bf u}{\bf v}=0\,, \qquad \overline{\bf v}{\bf u}=0\,.
\ee

The free solutions having the same sense of propagation 
are orthonormal over ${\mathbb R}^4$ in the continuum normalization, with  block form
\be\label{ORTHNCB}
\begin{aligned}\int d^4x\, \overline{{\bf f}^{(+)}_p(x,\tau)}{\bf f}^{(+)}_q(x,\tau)&=+\delta^4(p-q)I_2\,, \\  \int d^4x\, \overline{{\bf f}^{(-)}_p(x,\tau)}{\bf f}^{(-)}_q(x,\tau)&=-\delta^4(p-q)I_2\,,
\end{aligned}
\ee
while those having the opposite sense are orthogonal, with block form
\be\label{ORTHGCB}
\begin{aligned}
\int d^4x\, \overline{{\bf f}^{(+)}_p(x,\tau)}{\bf f}^{(-)}_q(x,\tau)&=0\,, \\ \int d^4x\, \overline{{\bf f}^{(-)}_p(x,\tau)}{\bf f}^{(+)}_q(x,\tau)&=0\,,
\end{aligned}
\ee
where ${\bf f}_p^{(\pm)}=(f_p^{(\pm)(1)},f_p^{(\pm)(2)})$ for $f^{(+)(r)}_p(x,\tau)=u^{(r)}(p)(1/2\pi)^2\exp[\rmi( p \cdot x+ \varphi_p m_p\tau)]$ and  $f^{(-)(r)}_p(x,\tau)=v^{(r)}(p)(1/2\pi)^2\exp[\rmi( p \cdot x- \varphi_p m_p\tau)]$, with $r=1,2$\,.  

The spin projection operators in an arbitrary frame are
\be\label{SPROJ}
P(\pm s)=\frac{1}{2}(I_4\mp \gamma^5\fy{s})\,.
\ee
In the rest frame, $s=(0,{\bf s})$ where $|{\bf s}|=1$\,. The operators  $\Lambda_u(p)\,,\Lambda_v(p)$ and $P(s)$ commute for all $s$ and $p$\,, since $\fy{p}\fy{s}=-p\cdot s=0$\,. 
The chirality projection operators are $P_\pm=(1 \pm \gamma^5)/2$, yielding for example the projections ${\bf u}_\pm=P_\pm{\bf u}$\,. The helicity operator for the unit momentum $\hat{\mathbf p}= {\mathbf p}/|{\mathbf p}|$ is
\be\label{hel}
\hat{\mathbf p}\cdot {\bf \Sigma}=\left(\begin{matrix}\hat{\mathbf p}\cdot\sigma & 0 \\ 0 & \hat{\mathbf p}\cdot\sigma \end{matrix}\right)\,.
\ee
In the ultrarelativistic limit as $m_p/E_p \to 0$, the projections ${\bf u}_{\pm}$ are eigenstates of the helicity operator with eigenvalues $\pm \varphi_p$ respectively. In the same limit, the projections ${\bf v}_\pm=P_\pm{\bf v}$ are helicity eigenstates also with eigenvalues $\pm\varphi_p$ respectively. 

Defining the free \wfs ${\bf h}_p^{(\pm)}$ by
\be\label{TPCF}
{\bf h}_p^{(\pm)}(x,\tau)=(\mathcal {TPC}\; {\bf f}_p^{(\mp)})(x,\tau)=-\rmi \gamma^5{\bf f}_p^{(\mp)}(-x,\tau)\,,
\ee
it follows that 


\be\label{HF}
{\bf f}_{-p}^{(\pm)}(x,\tau)=\rmi {\bf h}_{p}^{(\pm)}(x,\tau)\,.
\ee


Moreover,  the $\mathcal{TPC}$ conjugate of an eigenstate of the  forward projection $\Lambda_u(p)P(+s)$ is an eigenstate of $\Lambda_v(-p)P(-s)$\,. Thus, defining the antiparticle to be the  $\mathcal{TPC}$ conjugate of a particle, it follows that a positron is an electron having the reversed sense of propagation in time $t$ as $\tau$ increases, the reversed energy--momentum and spin,  but the  same coordinate velocity $dx^j/dt$ at constant phase and constant $\tau$\,. In particular, there are positrons of positive {\it and} negative energy. 

Again, particles and antiparticles may in general have energies of either sign, and may propagate forward or backward irrespective of the signs of their energies. The  space $\mathcal{S}$ of all subluminal states has the subspaces 
\begin{itemize}\label{subspaces}
\item[$\mathcal{S}_+$]  which  is spanned by particles  $f_p^{(+)(r)}(x,\tau)$ and antiparticles $h_p^{(-)(r)}(x,\tau)$, for $r=1,2$ and for all $p$ such that $ p^0>0$\,,
\end{itemize}
and 
\begin{itemize}
\item[$\mathcal{S}_-$]   which  is spanned by particles  $f_p^{(+)(r)}(x,\tau)$ and  antiparticles $h_p^{(-)(r)}(x,\tau)$, for $r=1,2$ and for all $p$ such that $ p^0<0$\,.
\end{itemize}
The two subspaces are a partition, that is, $\mathcal{S}_+\cap\mathcal{S}_-=\{0\}$ and $\mathcal{S}_+\cup\mathcal{S}_-=\mathcal{S}$\,. The case $p^0=0$ is precluded by admitting $m_p=0$ only as a limit. Note that the parenthetical superscripts $(\pm)$ on the free \wfs indicate the sense of propagation in time as $\tau$ increases. Also, the particles in $\mathcal{S}_+$ are all proportional to $\exp(+\rmi m_p\tau)$ , while those in $\mathcal{S}_-$ are all proportional to $\exp(-\rmi m_p\tau)$\,. Conservation of mass (see \S\ref{S:scatfirst})   prevents states in $\mathcal{S}_\pm$ from  being scattered into states in $\mathcal{S}_\mp$ by massless photons. It is however the case that states of  positive energy $p^0$ in $\mathcal{S}_+$, for example, can be scattered into $\mathcal{S}_+$ states of negative energy without lower bound and so hole theory must be invoked. Equivalently, the negative--energy electron states in $\mathcal{S}_+$ are regarded as states of different particles, namely positive--energy positrons \cite{Griff08}. Pair creation and annihilation are real processes, but the formalism of second quantization is avoided here by adopting the Feynman--Stueckelberg interpretation. That is, a positron in $\mathcal{S}_+$ is formally represented by  the $\mathcal{TPC}$--conjugate of an electron in $\mathcal{S}_+$\,.

\subsection{free \infls }\label{S:freeprop}
The two free \infls \; $\Gamma^0_\pm(x'-x,\tau'-\tau)$ for (\ref{PDIR}) both satisfy
\be\label{prop1}
\frac{1}{\rmi}\frac{\partial}{\partial \tau '}\Gamma^0_\pm(x'-x,\tau'-\tau)+\gamma_\mu\frac{1}{\rmi}
\frac{\partial}{\partial {x_\mu}' }\Gamma^0_\pm(x'-x,\tau'-\tau)=\delta^4(x'-x)\delta(\tau'-\tau)\,,
\ee
and are given by
\be\label{prop2}
\Gamma^0_\pm(x'-x,\tau'-\tau)=(2\pi)^{-5}\int dm\int d^4p\frac{mI_4-\fy{p}}{m^2-(m_p^2 \pm \rmi \epsilon)}\exp[\rmi(p\cdot (x'-x)+m(\tau'-\tau))]
\ee
where $0<\epsilon <<1 $. The small positive number $+\epsilon$ serves to remind that the inversion path  is below the  pole at $+m_p$ in the complex plane of $m$, and above the pole at $-m_p$\,.
Contour integration in the $m$--plane  yields
\be\label{prop3}
\Gamma^0_\pm(x'-x,\tau'-\tau)= 
\rmi \int d^4p\bigg\{\theta(\tau'-\tau)\mathcal{E}_\pm-\theta(\tau-\tau')\mathcal{E}_\mp\bigg\}\,,
\ee
where $\theta$ is again the Heaviside unit step function and
\be\label{prop4}
\mathcal{E}_\pm=(2\pi)^{-4}\bigg(\frac{\pm 
m_p I_4- \fy{p}}{\pm
2m_p}\bigg)\exp[\rmi(p\cdot (x'-x) \pm m_p(\tau'-\tau))]\,.
\ee
The two free \infls are related by
\be\label{prop4.5}
\overline{\Gamma^0_\pm(x'-x,\tau'-\tau)} \equiv \gamma^0\big(\Gamma^0_\pm(x'-x,\tau'-\tau)\big)^\dg\gamma^0=\Gamma^0_\mp(x-x',\tau-\tau')\,.
\ee

Considering $\Gamma^0_+$ for $\tau'-\tau>0$\,,
\be\label{prop5}
\Gamma^0_+(x'-x,\tau'-\tau)=  \rmi\int d^4p\bigg\{\theta(p^0){\bf f}_p^{(+)}(x',\tau')\overline {{\bf f}_p^{(+)}(x,\tau)} -\theta(-p^0){\bf f}_p^{(-)}(x',\tau')\overline {{\bf f}_p^{(-)}(x,\tau)}\bigg\}\,.
\ee
As $\tau'-\tau \to +\infty$,  $\Gamma^0_+$ evolves forward--propagating waves of positive energy and backward--propagating waves of negative energy. In terms of  the forward--propagating waves and their backward--propagating $\mathcal {TPC}$ conjugates waves,  $\Gamma^0_+$ becomes, for $\tau'-\tau>0$, 
\be\label{prop6}
\Gamma^0_+(x'-x,\tau'-\tau)=  \rmi\int d^4p\,   \theta(p^0)\bigg\{{\bf f}_p^{(+)}(x',\tau')\overline {{\bf f}_p^{(+)}(x,\tau)} -{\bf h}_p^{(-)}(x',\tau')\overline {{\bf h}_p^{(-)}(x,\tau)}\bigg\}\,.
\ee
Thus as $\tau'-\tau \to +\infty$, $\Gamma^0_+$ evolves only states in $\mathcal{S}_+$\,, that is the positive-energy, forward-propagating particles (for example, electrons) and the positive--energy, backward propagating antiparticles (positrons).

Considering $\Gamma^0_+$ for $\tau'-\tau<0$\,,
\be\label{prop6.1}
\Gamma^0_+(x'-x,\tau'-\tau)=  \\ \rmi\int d^4p\bigg\{\theta(p^0){\bf f}_p^{(-)}(x',\tau')\overline {{\bf f}_p^{(-)}(x,\tau)} -\theta(-p^0){\bf f}_p^{(+)}(x',\tau')\overline {{\bf f}_p^{(+)}(x,\tau)}\bigg\}\,.
\ee
 
 Thus as $\tau'-\tau \to -\infty$,  $\Gamma^0_+$ evolves  backward--propagating \wfs of positive energy and forward--propagating \wfs of negative energy. In terms of  the forward--propagating \wfs and their backward-propagating  $\mathcal {TPC}$ conjugate wave functions,  $\Gamma^0_+$ becomes, for $\tau'-\tau<0$, 
\be\label{prop6.2}
\Gamma^0_+(x'-x,\tau'-\tau)=  \\- \rmi\int d^4p\,   \theta(-p^0)\bigg\{{\bf f}_p^{(+)}(x',\tau')\overline {{\bf f}_p^{(+)}(x,\tau)} -{\bf h}_p^{(-)}(x',\tau')\overline {{\bf h}_p^{(-)}(x,\tau)}\bigg\}\,.
\ee
Thus as $\tau'-\tau \to -\infty$, $\Gamma^0_+$ evolves only states in $\mathcal {S}_-$\,, that is, the negative--energy, forward--propagating particles and the negative--energy, backward--propagating antiparticles. Similarly it follows that for $\tau'-\tau>0$\,,
\be\label{prop6.3}
\Gamma^0_-(x'-x,\tau'-\tau)=  \\ \rmi\int d^4p\,   \theta(-p^0)\bigg\{{\bf f}_p^{(+)}(x',\tau')\overline {{\bf f}_p^{(+)}(x,\tau)} -{\bf h}_p^{(-)}(x',\tau')\overline {{\bf h}_p^{(-)}(x,\tau)}\bigg\}\,,
\ee
while for $\tau'-\tau<0$\,,
\be\label{prop6.2-}
\Gamma^0_-(x'-x,\tau'-\tau)=  \\ -\rmi\int d^4p\,   \theta(p^0)\bigg\{{\bf f}_p^{(+)}(x',\tau')\overline {{\bf f}_p^{(+)}(x,\tau)} -{\bf h}_p^{(-)}(x',\tau')\overline {{\bf h}_p^{(-)}(x,\tau)}\bigg\}\,.
\ee
It is readily seen that
\be\label{prop6.4}
\Gamma^0_\pm(x'-x,\tau'-\tau)=\mathrm {sign}(\tau'-\tau)\frac{1}{\rmi}\int d^4x''\,\Gamma^0_\pm(x'-x'',\tau-\tau'')\Gamma^0_\pm(x''-x,\tau''-\tau)
\ee
for $\tau' > \tau'' >\tau$ in which case  $\mathrm {sign}(\tau'-\tau)=+1$, and also for $\tau' <\tau''<\tau$ in which case  $\mathrm {sign}(\tau'-\tau)=-1$. If either $\tau'' < \tau$ or $\tau''> \tau'$, then the right hand side of (\ref{prop6.4}) vanishes.

\subsection{M{\o}ller operators}\label{S:IF}
The \infls $\Gamma_\pm(x',\tau';x,\tau)$ are defined as solutions of 
\be\label{IF1}
\frac{1}{\rmi}\frac{\partial}{\partial \tau'}\Gamma_\pm(x',\tau';x,\tau)+\gamma_\mu\Big( \frac{1}{\rmi}
\frac{\partial}{\partial {x_\mu} '}-e A^\mu (x')\Big)\Gamma_\pm(x',\tau';x,\tau) \\ =\delta^4(x'-x)\delta(\tau'-\tau)\,.
\ee
They depend upon the charge $e$ and potential $A^\mu$, but  in the interest of clarity the argument lists here for $\Gamma_\pm$ include only the events $x,x'$ and parameters $\tau,\tau'$.  The boundary conditions are
\be\label{MoBC+}
\Gamma_+(x',\tau';x,\tau) \sim \Gamma_+^0(x'-x,\tau'-\tau)
\ee
as $\tau',\tau \to -\infty$ with $\tau'>\tau$ and
\be\label{MoBC-}
\Gamma_-(x',\tau';x,\tau) \sim \Gamma_-^0(x'-x,\tau'-\tau)
\ee
as $\tau',\tau \to +\infty$ with $\tau>\tau'$\,. 
Rearranging (\ref{IF1})  yields \cite{BjDr64}
\be\label{IF2}
\Gamma_\pm(x',\tau';x,\tau)=\Gamma^0_\pm(x'-x,\tau'-\tau) \\+\int d\tau'' \int d^4x'' \,\Gamma^0_\pm(x'-x'',\tau'-\tau'')e\fy{A}(x'')\Gamma_\pm(x'',\tau'';x,\tau)\,.
\ee
Expanding (\ref{IF2}) in the Born series shows that 
\be\label{IF3}
\overline{\Gamma_\pm(x',\tau';x,\tau)}=\Gamma_\mp(x,\tau;x',\tau')\,.
\ee
The semigroup property (\ref{prop6.4}) holds also for $\Gamma_\pm$.

For $\phi_i  \in \mathcal{S_+}$\,, define $\omega_+\phi_i$ by
\be\label{IF4}
\omega_+\phi_i(x',\tau')=\lim_{\tau \to -\infty}\frac{+1}{\rmi}\int d^4x\,\Gamma_+(x',\tau';x,\tau)\phi_i(x,\tau)\,.
\ee
If the support of the potential $A^\mu$ is bounded in ${\mathbb E^4}$, and if the support of $\phi_i(x,\tau) $ is exterior to that of $A^\mu$ as $\tau \to \pm \infty$, then  $\omega_+\phi_i$ satisfies (\ref{PDIR}) subject to $\omega_+\phi_i(x',\tau') \to \phi_i(x',\tau')$ as $\tau' \to - \infty$\,. The \wf  $\omega_+\phi_i$ may be expressed as  
\be\label{prop8}
\omega_+\phi_i(x',\tau') = \phi_i(x',\tau') + \int d^4x \int^{+\infty}_{-\infty} d\tau\,\Gamma^0_+(x'-x,\tau'-\tau) e \fy{A}(x)\omega_+\phi_i(x,\tau)\,.
\ee
It may be seen from (\ref{prop8}) that $\omega_+\phi_i(x,\tau) - \phi_i(x,\tau) \in \mathcal{S}_\pm$ as $\tau \to \pm\infty$, that is, there is `$\tau$--increasing' scattering into positive--energy, forward--propagating particles and positive--energy, backward--propagating antiparticles, and also  `$\tau$--decreasing' scattering  into negative-energy, forward--propagating particles and negative--energy, backward--propagating antiparticles. If $\phi_i \in \mathcal{S}_-$ then $\omega_+\phi_i=0$.

Similarly for $\phi_f \in \mathcal{S}_+$\, (not $\mathcal{S}_-$) define $\omega_-\phi_f(x,\tau)$ by  
\be\label{IF4-}
\omega_-\phi_f(x',\tau')=\lim_{\tau \to +\infty}\frac{-1}{\rmi}\int d^4x\,\Gamma_-(x',\tau';x,\tau)\phi_f(x,\tau)\,.
\ee
Then $\omega_-\phi_f$ satisfies (\ref{PDIR}), subject to $\omega_-\phi_f(x',\tau') \to \phi_f(x',\tau')$ as $\tau' \to +\infty$ and  
\be\label{prop9}
\omega_-\phi_f(x',\tau') = \phi_f(x',\tau') + \int d^4x \int^{+\infty}_{-\infty} d\tau\,\Gamma^0_-(x'-x,\tau'-\tau) e \fy{A}(x)\omega_-\phi_f(x,\tau)\,.
\ee
It may be seen from (\ref{prop9}) that $\omega_-\phi_f(x,\tau) - \phi_f(x,\tau) \in \mathcal{S}_\mp$ as $\tau \to \pm\infty$\,. If $\phi_f \in \mathcal{S}_-$ then $\omega_-\phi_f=0$\,.

In summary, the M{\o}ller operators $\omega_{\pm}$ are defined by
\be\label{Mpm}
\omega_{\pm}\phi(x',\tau')=\lim_{\tau \to \mp \infty} \frac{\pm 1}{\rmi}\int d^4x\,\Gamma_{\pm}(x',\tau';x,\tau)\phi(x,\tau)
\ee
for all $\phi \in \mathcal{S_+}$\,. Again, $\omega_{\pm}\phi =0$ for all  $\phi \in \mathcal{S}_-$\,.

The potential $A^\mu$ involved in the M\o ller operators is calculated as the particular integral of (\ref{MAX}) for the concatenation of the M{\o}ller current
\be\label{Mol4}
j_{fi}^\mu(x,\tau)=\overline {\omega_-\phi_f(x,\tau)}\gamma^\mu\omega_+\phi_i(x,\tau)=\overline {\phi_f(x,\tau)}\gamma^\mu\phi_i(x,\tau)+{\mathcal O}(e^2)\,,
\ee
assuming that $A^\mu = {\mathcal O}(e)$\,.

\section{Two Dirac particles}\label{S:2part}
The single--particle formalism of the preceding sections is extended to two particles. Space--time entanglement  is displayed in an elementary and unrestricted way. A proof of the spin--statistics theorem is very briefly described, and the Bethe--Salpeter equation for bound states is derived from the two--particle formalism without further conjecture.  The mutual scattering problem for two particles is reduced to that of a single particle in the presence of an `external' potential.
 
\subsection{two--particle wave equation}\label{S:2partwave}

The single--particle, spin--1/2  wave equation (\ref{PDIR}) and Lagrangian (\ref{LAG}) are readily extended to two--particle, spin--1/2 \wfs in the  tensor product space. The wave equation becomes

\be\label{TWO}
\frac{1}{\rmi}\frac{\partial}{\partial \tau}\Psi+\fy{\pi}(x)\otimes I_4\,\Psi+I_4\otimes \fy{\pi}(y)\,\Psi=0
\ee
where $\pi^\mu(x)=(1/\rmi)\partial /\partial x_\mu-e_1A^\mu(x)$ and $\pi^\mu(y)
=(1/\rmi)\partial /\partial y_\mu-e_2A^\mu(y)$\,. 

Consider first the case of externally-sourced electromagnetic potentials $A^\mu(x)$ and $A^\mu(y)$\,. Two indistinguishable spin--1/2 particles entangled at the events $x=(t,\bf{x})$ and $y=(s,\bf{y})$  may be represented in the standard way with antisymmetric combinations of tensor products of single--particle 4--spinor wavefunctions, such as  
\be\label{fermions}
\Psi(x,y,\tau)=\frac{1}{\sqrt {2}}\Big(\psi(x,\tau)\otimes\chi(y,\tau)-\chi(x,\tau)\otimes\psi(y,\tau)\Big)\,.
\ee
 This representation of entanglement, in space or time, is simpler and more general than the QFT representations. 

Consider now interaction potentials and self potentials acting on both particles. The potentials are constructed by integrating over the dependences of the two--particle currents upon one of the particles. That is,
\be\label{MaxTwo}
\square A^\mu(x)=e_1 J_1^\mu(x)+e_2 J_2^\mu(x)
\ee
where $e_1$ and $e_2$ are the charges of the two particles. The currents are, after concatenation and marginalization, 

\be\label{CatTwo1}
J_1^\mu(x)=\int d \tau \int d^4y\; \overline{\Psi(x,y,\tau)}\,\gamma^\mu\otimes I_4\,\Psi(x,y,\tau)
\ee
and
\be\label{CatTwo2}
J_2^\mu(x)=\int d\tau\int d^4y\; \overline{\Psi(y,x,\tau)}\, I_4\otimes\gamma^\mu\,\Psi(y,x,\tau)\,.
\ee
In particular, a particle may experience the potential arising from its own current.  If the two particles are indistinguishable, the \wf $\Psi$ is chosen to be antisymmetric as in (\ref{fermions})\,. Then the corresponding quantum currents  $J_1^\mu, J_2^\mu$ and hence the semiclassical field $A^\mu$ are unaltered by permutation of the two single--particle states at the same parameter $\tau$. That is, the currents and the potential  are appropriately bosonic. 

If the particles are distinguishable and the \wf is a simple tensor product $\Psi(x,y,\tau)=\psi(x,\tau)\otimes\chi(y,\tau)$, where both $\psi$ and $\chi$ are normalized beams or packets, then the concatenated currents reduce to 
\be\label{Beam1}
J_1^\mu(x)=\int d\tau\,\overline{\psi(x,\tau)}\,\gamma^\mu\psi(x,\tau)
\ee
and
\be\label{Beam2}
J_2^\mu(x)=\int d\tau\, \overline{\chi(x,\tau)}\,\gamma^\mu\chi(x,\tau)\,.
\ee

\subsection{spin and statistics}\label{S:spinstat}

The spin--statistics connection is usually presented \cite{ Wei95, Zee03, IZ, Sr07} as a theorem in relativistic QFT. On the other hand, the unimodular and unitary group $SU(2)$ spanned by the Pauli spin matrices is a spin--1/2 representation of  the unimodular and orthogonal group $SO(3)$ of spatial rotations \cite{WeiQM13}. That is, spin is not intrinsically a relativistic phenomenon. Indeed, the single--particle, spin--1/2 Pauli wave equation \cite{BjDr64} is covariant with respect to $SO(3)$ provided that the electric potential ${\bf A}$ and the magnetic field $\bf{B}$ are functions of $|\bf{x}|$ alone. The spin--statistics connection has furthermore been proved for both nonrelativistic and relativistic quantum mechanics  of arbitrary spin \cite{Jabs10}.  The proof takes into account the phases  of spin eigenstates. The phases are indeterminate since only a spin axis is specified, rather than a spin frame.  Homotopically consistent permutations of the arbitrary phases between numerous otherwise indistinguishable particles, along with permutations of the conventionally observable positions and  spins, lead to the standard  spin--statistics connection. The proof holds in particular for relativistic spin--1/2 quantum mechanics. The phase  is not a standard  observable, but  the universal observance of the exclusion principle    impresses physical significance upon the phase indeterminacy.

\subsection{two--particle M\o ller operators}\label{S:2PROP}

There are four  two--particle free influence functions.  Of particular interest are $\Gamma^0_{++}$ and $\Gamma^0_{--}$ given by 
\be\label{2ppg}
\Gamma^0_{\pm\pm}(x'-x,y'-y,\tau'-\tau)=\frac{1}{\rmi}\Gamma^0_{\pm}(x'-x,\tau'-\tau)\otimes \Gamma^0_{\pm}(y'-y,\tau'-\tau)\,.
\ee
Proceeding as in Section \ref{S:IF} leads to the two--particle M\o ller operators $\Omega_{++}$ and $\Omega _{--}$ acting on $\mathcal{S}_+\otimes \mathcal{S}_+$, such that 

\be\label{Mpm2}
\Omega_{\pm \pm}\Phi(x',y',\tau')=\lim_{\tau \to \mp \infty} \frac{\pm 1}{\rmi}\int d^4x \int d^4 y \,\Gamma_{\pm \pm}(x',y',\tau'; x,y,\tau)\Phi(x,y,\tau)
\ee
for all $\Phi \in \mathcal{S}_+\otimes \mathcal{S}_+$\,. The dependence of the influence functions $\Gamma_{\pm\pm}$ and the M\o ller operators $\Omega_{\pm\pm}$ upon the two charges $e_1$ and $e_2$,  and upon the potential  $A^\mu$ at $x$ and at $y$,  is of fundamental importance but for clarity is not made explicit here. It follows that $\Omega_{\pm \pm}\Phi =0$ for all  $\Phi \in \mathcal{S}\otimes\mathcal{S} \setminus \mathcal{S}_+\otimes \mathcal{S}_+$\,. The \infls $\Gamma_{\pm \pm}$ satisfy
\begin{multline}\label{2IF2}
\Gamma_{\pm \pm}(x',y',\tau';x,y,\tau)=\Gamma^0_{\pm \pm}(x'-x,y'-y,\tau'-\tau) \\+\int d\tau'' \int d^4x'' \int d^4y''\,\Gamma^0_{\pm \pm}(x'-x'',y-y'',\tau'-\tau'')\\ \times V(x'',y'')\Gamma_{\pm \pm}(x'',y'',\tau'';x,y,\tau)
\end{multline}
The scattering potential in (\ref{2IF2}) is 
\be\label{intpot2}
V(x,y)=
\left(
\begin{matrix}
e_1\fy{A}(x) & 0 \\ 
0 &e_2\fy{A}(y)
\end{matrix}
\right)\,.
\ee
The semigroup property (\ref{prop6.4}) extends to $\Gamma_{\pm \pm}$\,. When the potential $A^\mu$ is external, it may be shown that
\be\label{sep2}
\Gamma_{\pm \pm}(x',y',\tau';x,y,\tau)=\frac{1}{\rmi}\Gamma_{\pm}(x',\tau';x,\tau)\otimes\Gamma_{\pm}(y',\tau';y,\tau)\,,
\ee
and that
\be\label{hat2}
\overline{\Gamma_{\pm \pm}(x',y',\tau';x,y,\tau)}=\Gamma_{\mp \mp}(x,y,\tau;x',y',\tau')\,.
\ee
The separability property (\ref{sep2}) implies that unentanglement is conserved with the passage of $\tau$. The M\o ller currents are
 \be\label{CatTwo1m}
J_1^\mu(x)_{fi}=\int d \tau \int d^4y\; \overline{\Omega_{--}\Phi_f(x,y,\tau)}\,\gamma^\mu\otimes I_4\,\Omega_{++}\Phi_i(x,y,\tau)
\ee
and
\be\label{CatTwo2m}
J_2^\mu(x)_{fi}=\int d\tau\int d^4y\; \overline{\Omega_{--}\Phi_f(y,x,\tau)}\, I_4\otimes\gamma^\mu\,\Omega_{++}\Phi_i(y,x,\tau)\,.
\ee

\subsection{bound states}
It is now convenient to combine the \infls in the form
\be\label{onegam}
\Gamma(x',y',\tau';x,y,\tau)=\theta(\tau'-\tau)\Gamma_{++}(x',y',\tau';x,y,\tau)\\-\theta(\tau-\tau')\Gamma_{--}(x',y',\tau';x,y,\tau)\,,
\ee
and similarly there is the combined free \infl $\Gamma^0(x'-x,y'-y,\tau'-\tau)$\,. Then for all $\tau'$ and $\tau$\,, any two--particle \wf $\Psi(x,y,\tau)$ evolves as
\be\label{wfevol}
\Psi(x',y',\tau')=\frac{1}{\rmi}\int d^4x \int d^4y\, \Gamma(x',y',\tau';x,y,\tau)\Psi(x,y,\tau)\,,
\ee

 or by virtue of (\ref{2IF2}) as
\begin{multline}\label{Psiprime}
\Psi(x',y',\tau')=\Psi_0(x',y',\tau') \\+\int d\tau'' \int d^4x'' \int d^4y''\,\Gamma^0(x'-x'',y-y'',\tau'-\tau'')V(x'',y'')\Psi(x'',y'',\tau'')
\end{multline}
where the freely propagated \wf $\Psi_0(x',y',\tau')$ is 

\be\label{freePsi}
\Psi_0(x',y',\tau')=\frac{1}{\rmi}\int d^4x \int d^4 y\, \Gamma^0(x'-x,y'-y,\tau'-\tau)\Psi(x,y,\tau)\,.
\ee

The Fourier transform of (\ref{Psiprime}) with respect to $\tau'$ is the inhomogeneous Bethe--Salpeter equation \cite{Bethe51,Salp51,Grei03}
\begin{multline}\label{FouPsi}
\Psi(x',y',m)=\Psi_0(x',y',m) \\+\int d^4x'' \int d^4y''\,\Gamma^0(x'-x'',y-y'',m)V(x'',y'')\Psi(x'',y'',m)\,,
\end{multline}
in the casual notation where
\be\label{casual}
f(m)=\int d \tau\,f(\tau)\exp(-\rmi m \tau)\,.
\ee 
The interaction potential $V(x,y)$ as given in (\ref{intpot2}) is determined semi-classically, that is, in terms of potentials $A^\mu(x)$  satisfying Maxwell's equation (\ref{MaxTwo}) for the Dirac currents (\ref{CatTwo1}) and (\ref{CatTwo2}). Bound states are defined as eigenstates of the homogeneous equation \cite{Grei03}, that is, (\ref{FouPsi}) for $\Psi_0=0$\,. Indeed, if there is binding energy in the two--particle state $\Psi$ then the freely--propagating $\Psi_0$ is a kinematical impossibility \cite{Salp51, Grei03}. The Bethe--Salpeter equation is in general nonlinear but may be expanded at least formally in powers of the charges $e_1$ and $e_2$\,, yielding a series of linear equations. A method of partial summation is described in \S\ref{S:oneloop} below. 
 
\subsection{two--particle scattering matrix}\label{S:per}
Let $\Phi_i$ and $\Phi_f$ be two free \wfs in $\mathcal{S}_+\otimes \mathcal{S}_+$\,. Then the scattering matrix $S_{fi}$ is defined by
\be\label{sm1}
S_{fi} \equiv \int d^4x \int d^4y\; \overline{\Omega_{--}\Phi_f(x,y,\tau)}\Omega_{++}\Phi_i(x,y,\tau)\,.
\ee
The incident and final two--particle states for indistinguishable particles are independently fermionic. That is, the incident single--particle states may be permuted independently of the final states and vice versa, leading in each case to a reversal of the sign of $S_{fi}$\,.

It is readily shown that $\partial S_{fi}/\partial \tau =0$\,, since $\fy{\pi}^\dg\gamma^0=\gamma^0\fy{\pi}$\,. Alternatively, it may be shown that 
\be\label{sm2}
S_{fi}=\lim_{\tau \to + \infty}\int d^4x \int d^4y\; \overline{\Phi_f(x,y,\tau)}\Omega_{++}\Phi_i(x,y,\tau)\,.
\ee
which is explicitly independent of $\tau$. It follows from (\ref{sm2}) that
\be\label{sm3}
S_{fi}=S^{(0)}_{fi}+\rmi \int d\tau \int d^4x \int d^4y\; \overline{\Phi_f(x,y,\tau)}V(x,y)\Omega_{++}\Phi_i(x,y,\tau)\,,
\ee
where
\be\label{sm4}
S^{(0)}_{fi} = \int d^4x \int d^4y\; \overline{\Phi_f(x,y,\tau)}\Phi_i(x,y,\tau)
\ee
which is independent of $\tau$ assuming that  $\Phi_i$  and $\Phi_f$ are free particle pairs. If the particles are distinguishable then the incident and final free \wfs are of the form $\Phi_i=\phi_i\otimes\xi_i$ and $\Phi_f=\phi_f\otimes\xi_f$ respectively, while the M\o ller operators are of the form $\Omega_{\pm\pm}=\omega_\pm\otimes\omega_\pm$. The M\o ller currents are calculated, following (\ref{Beam1}) and (\ref{Beam2}), as
\be\label{Mo1}
J_1^\mu(x)=\int d \tau \,  \overline{\omega_-\phi_f(x,\tau)}\,\gamma^\mu \omega_+\phi_i(x,\tau)
\ee
and
\be\label{Mo2}
J_2^\mu(x)=\int d\tau\,  \overline{\omega_-\xi_f(x,\tau)}\,\gamma^\mu \omega_+\xi_i(x,\tau)\,.
\ee
If the two particles are indistinguishable then  both the incident and final two--particle free states $\Phi_i$  and  $\Phi_f$  in (\ref{sm1}) must be fermionic as in (\ref{fermions}). However,  both two--particle free states in the M\o ller currents (\ref{CatTwo1m}) and (\ref{CatTwo2m}) must be replaced with  bosonic states of the form
\be\label{boson}
\Phi(x,y,\tau)=\frac{1}{\sqrt{2}}\Big(\phi(x,\tau)\otimes\xi(y,\tau)+\xi(x,\tau)\otimes\phi(y,\tau)\Big)\,.
\ee
The M\o ller currents  are then bosonic,  prior to marginalization. It follows eventually that the scattering matrix is appropriately fermionic with respect to the incident two--particle state, and also with respect to the final two--particle state.

The integrand for nontrivial scattering in (\ref{sm3}) is, in the case of distinguishable particles for simplicity,
\begin{multline}\label{nontriv}
 \overline{\Phi_f(x,y,\tau)}V(x,y)\Omega_{++}\Phi_i(x,y,\tau)= \\ \overline{\phi_f(x,\tau)}e_1\fy{A}(x)\omega_+\phi_i(x,\tau)\overline{ \xi_f(y,\tau)}\omega_+\xi_i(y,\tau) \\ + \overline{\phi_f(x,\tau)}\omega_+\phi_i(x,\tau)\overline{ \xi_f(y,\tau)}e_2\fy{A}(y)\omega_+\xi_i(y,\tau)\,.
 \end{multline}
Iteration on (\ref{sm4}) by expansion in powers of the charges $e_1$ and $e_2$ proceeds the most efficiently if the M\o ller operators in the factors not explicitly displaying scattering, that is in the factors $\overline{ \xi_f}\omega_+\xi_i$ and $\overline{\phi_f}\omega_+\phi_i$,  are not so iterated. These factors reduce to unity, after forming the scattering cross--section and averaging over the incident and final states. It suffices therefore to consider single--particle scattering off the semiclassical potential owing to any  source including the current of the particle itself. The incident and final energy--momenta $p_i^\mu$ and $p_f^\mu$ are assumed timelike, hence the ostensibly `external' photon is spacelike.

The potential owing to an `external' source ${\mathcal Z}^\mu(x)$ is now denoted ${\mathcal A}^\mu(x)$, with
\be\label{MAXEXT}
\square \mathcal{A}^\mu =\mathcal{Z}^\mu\,.
\ee
If the external source $\mathcal{Z}^\mu$ is in fact the current $e_2J_2^{\;\mu}$ of a second  and  possibly distinguishable spin--1/2 particle then, owing to the parity symmetry of the D'Alembertian $\square$\,, it is readily shown that the single--particle scattering matrix is symmetric with respect to the two particles. That is, the scattering of the second particle by the first is the same as the scattering of the first particle by the second.  

\section{First--order scattering of a single particle} \label{S:scatfirst}
The scattering matrix for a single particle resembles (\ref{sm1}) or equivalently  (\ref{sm3}). To leading order in powers of the charge $e$\,, the nontrivial contribution to the analog of (\ref{sm3}) is
\be\label{first1}
S^{(1)}_{fi}=\rmi\int d\tau \int d^4x \,\overline{\phi_f(x,\tau)}\,e\fy{\mathcal A}(x)\phi_i(x,\tau)\,.
\ee
Assume that both $\phi_i$ and $\phi_f$ have the form (\ref{FWD}) with both $p^0_i$ and $p^0_f$ positive, hence both  \wfs are in $\mathcal{S}_+$. It follows that 
\be\label{first2}
S^{(1)}_{fi}=\frac{\rmi e}{(2\pi)^3}\delta(\Delta m_p)\, \overline{u_f} \, \fy{\mathcal{A}}(\Delta p) \, u_i 
\ee
where $\Delta p = p_f-p_i$ and $\Delta m_p=m(p_f)-m(p_i)$, while  $u_i={\bf u}(p_i){\bf a}_i$ and  $u_f={\bf u}(p_f){\bf a}_f$. The coefficients ${\bf a}_i$ and ${\bf a}_f$ are complex $2 \times 1$  matrices. Finally, $\mathcal{A}^\mu(\Delta p)$ is the Fourier transform of $\mathcal{A}^\mu(x)$ defined by
\be\label{ Ffwd}
\mathcal{A}^\mu(\Delta p)=\int d^4x \,\mathcal{A}^\mu(x)\exp[-\rmi\Delta p \cdot x]\,.
\ee
Note that the mass of the particle is conserved, that is, $m_f=m(p_f)=m(p_i)=m_i$ which may as well be assigned the notation $m_e$ for, say, an electron.

The  Coulomb potential for  a point charge $-Ze >0$ is defined by $\mathcal{A}^0(x)=-Ze/(4\pi |\mathbf{x}|)$ where $x=(x^0,\mathbf{x})$, and by $\mathcal{A}^j(x)=0$ for $j=1,2,3$\,. Hence $A^0(\Delta p)= -2\pi Ze\,\delta(\Delta p^0)/|\Delta \mathbf{p}|^2$\,. Conservation of energy $p^0$ as well as mass $m(p)$  implies that the magnitude  of momentum $\mathbf{p}$ is also conserved, and so a scattering parameter $\kappa$ may be defined  by ${\mathbf p}_f\cdot {\mathbf p}_i=|\mathbf{p}|^2 \cos \kappa$\,. It follows from (\ref{first2}) in the standard way \cite{BjDr64, MandM65} that the effect of spin is to modify the $\kappa$--dependence  of the Rutherford cross section by the Mott factor ${\mathcal M}=1-|\mathbf{p}/m_e|^2\sin^2(\kappa/2)$\,.

It may be remarked in passing that, for the purposes of calculating cross--sections, the continuum normalizations (\ref{FWD}) and (\ref{BKD}) are  replaced with 4--cube normalizations rather than the standard 
 3--cube. Specifically, $(2\pi)^4$ is replaced with $L^4$ where $L$ is the edge of the 4--cube. The cross--section as calculated is then a volume rather than the standard area. In the case of Coulomb scattering, where the potential is independent of coordinate time $x^0$, the cross-sectional volume is proportional to $T_0 = 2\pi\delta(p^0_i-p^0_i)$ which is interpreted in the standard way \cite{BjDr64} as a large interval in $x^0$ during which scattering is observed. The incident \wf is in practice a beam having finite energy spread. The beam accordingly takes a finite time to sweep through an external scattering potential, although it must be conceded that such a time is undefined for an inverse--distance potential.  Thus it is in general appropriate when the potential is time--independent to divide the cross--sectional volume by $T_0$, yielding a cross--sectional area \cite{HoLa82}.

\section{One--loop corrections}\label{S:oneloop}
Assuming as in Coulomb scattering that $\mathcal{A}$ is ${\mathcal O}(e)$, the first--order external scattering $S^{(1)}_{fi}$ given by (\ref{first1}) is ${\mathcal O}(e^2)$\,. One--loop corrections $S^{(1)}_{fi}$ arise  from (i) self--scattering corrections to (\ref{first1}) and  (ii) external--scattering corrections to first--order self scattering. There are two contributions to (i), obtained by iteration on (\ref{IF2}) yielding $\Gamma_+$ correct to ${\mathcal O}(e^2)$\,. There are also two contributions to (ii), obtained by iteration on (\ref{Mpm}) yielding $\omega_\pm$ correct to ${\mathcal O}(e^2)$\,. The contributions to (ii) are, by virtue of (\ref{prop4.5}), identical to those to (i). Hence  there are in all four relevant corrections of  ${\mathcal O}(e^4)$ with total $M_4=2M_{12} + 2M_{21}$ where
\be\label{M12}
M_{12}=\rmi \int d\tau \int d^4x\,\int d\tau' \int d^4x'\;
 \overline{\phi_f(x,\tau)}\,e\, \fy{\mathcal{A}}(x)\Gamma^0_+(x-x',\tau-\tau')\,e\,\fy{A}(x')\phi_i(x',\tau')\,,
\ee
while $M_{21}$ differs from $M_{12}$ only by exchanging the potentials $\fy{\mathcal{A}}$ and $\fy{A}$ in (\ref{M12}). The self potential $A^\mu$ in (\ref{M12}) satisfies
\be\label{maxself}
\square A^\mu(x) = e J^\mu(x) = e\int d\tau \,\overline{\phi_f(x,\tau)}\,\gamma^\mu\phi_i(x,\tau)\,,
\ee
and is obtained as 
\be\label{solmax}
A^\mu(x)=e\int d^4x'\,D_F(x-x')J^\mu(x')
\ee
where the Feynman photon propagator $D_F$ is \cite{BjDr64}
\be\label{scam6}
D_F(x-x')=\int \frac{d^4k}{(2\pi)^4}\frac{1}{k^2-\rmi \epsilon}\exp[\rmi k\cdot(x-x')]\,.
\ee
Again, the metric here has the signature $(-+++)$.  

A change to the ${\mathcal O}(e^4)$ scattering matrix contribution $M_4=M_{12}+M_{21}$ is now made, by the substitution 
\be\label{Gapprox}
\phi^{\,}_i(x,\tau)\, \overline{\phi_f(x',\tau')} \Rightarrow\;  \\ \frac{1}{\rmi T_\tau} \, \exp[\rmi( m_i\tau-m_f\tau')]\int\frac{d^4r}{(2\pi)^4}\frac{\overline{m}I_4-\fy{r}}{\overline{m}^2+r^2-\rmi \epsilon}\exp[\rmi r \cdot(x-x')]
\ee
where $\overline{m}=(m_f+m_i)/2$, but the substitution (\ref{Gapprox}) is made 
{\it without} regard to the $\tau$--ordering.  The standard additional minus sign is introduced for a closed electron loop \cite{BjDr64}. The substitution is accurate for incident and final free--\wf beams $\phi_i$ and $\phi_f$, if the necessarily spacelike impact parameter $\Delta p =p_f-p_i$ for the beam axes is small. The scale $T_\tau = 2\pi\delta(m_i-m_i)$, for the parameter  $\tau$,  is  the inverse of the small mass spread of the  beams. In the far field the beams consist of free particles all on the same mass shell $m_i=m_f=m_e$, but the scattering takes place off mass shell and so an ${\mathcal O}(m_ee^2)$ mass spread is defined for all internal fermion lines.    The quantum--mechanical substitution (\ref{Gapprox}) has an analog in QFT \cite{Wei95,Zee03}, where the free fermion propagator is a two--event correlation. Making the  substitution (\ref{Gapprox}) in the expansion of the Bethe--Salpeter equation (\ref{FouPsi}) yields all of the standard diagrams \cite{Grei03}. 

The substitution (\ref{Gapprox}) having been made in the two copies of the scattering matrices $M_{12}$ and $M_{21}$ in all four topologically different ways, it is the case that the vertex  is modified by a factor $\mathcal{F}$ that is a sum of the four standard Feynman diagrams in Fig. 8--10 (b), (c), (d) and (e) of Bjorken and Drell \cite {BjDr64}.  In order of renormalization the diagrams are the  vacuum polarization of the external field (e),   the mass--renormalization counter term (d) and self--mass (c) on both the incident  and final fermion lines, and the vertex correction (b). The four diagrams are different, even though they arise here from two duplicated integrals, hence the symmetry factor for each diagram is $S=1$. Each of the four additive contributions to  the modification  factor $\mathcal{F}$ is independent of the parameter scale $T_\tau$\,.  The contributions are, relative to $S^{(1)}_{fi}$, precisely the standard wavenumber integrals in Eqs (8.8), (8.34) and (8.49) of Bjorken and Drell \cite{BjDr64}.  All the integrals derived here have the signs as deduced from QFT \cite{Zee03, IZ, Sr07}. That is, there is no need to introduce a sign correction by fiat, as is done by  Bjorken and Drell \cite{BjDr64} who assume the connection between spin and statistics. It is concluded that the Uehling potential, the anomalous magnetic moment of the electron and the Lamb shift have the respective leading--order values as eventually deduced by Bjorken and Drell \cite{BjDr64} and as accurately observed \cite{BjDr64}. 

The above--mentioned diagrams, integrals  and resulting one--loop corrections  are obtained by Bjorken and Drell \cite{BjDr64} from the standard Dirac wave equation for a single particle.  Instead of making the substitution (\ref{Gapprox}),  those authors replace the potential product $A^\mu(x)A^\nu(y)$ in the second--order scattering matrix with a superposition of products of leading--order M\o ller currents. The resulting matrix is then symmetric in its dependence upon the two particles. Bjorken and Drell refer  to their text on QFT \cite{BjDr65} for strict justification. 

\section{Axial anomaly}\label{S:axanom}
The parametrized Dirac equation (\ref{PDIR}) leads to the vector current $j^\mu=\overline{\psi}\gamma^\mu\psi$ obeying the identity (\ref{VCUR}), and 
also  to the axial current $j^\mu_5=\overline{\psi}\gamma^\mu\gamma_5\psi$ obeying
\be\label{axi}
 \overline{\psi}\gamma_5\frac{\partial \psi} {\partial \tau}-\frac{\partial \overline{\psi}}{\partial \tau}\gamma_5\psi+
 \frac{\partial}{\partial x^\mu}\overline{\psi}\gamma^\mu\gamma_5\psi=0\,.
 \ee
 The concatenated vector current $J^\mu$ is again divergenceless, that is, 
 \be\label{divvec}
 \partial_\mu J^\mu=\int j^\mu d\tau=0\,,
 \ee
while for the sharp mass  value $m_e$ the divergence of the concatenated axial current $J^\mu_5$ is
\be\label{divaxi}
\partial_\mu J^\mu_5 =\int j^\mu_5 d\tau=-2\rmi m_e \int \overline {\psi}\gamma_5 \psi d \tau \,.
\ee
The preceding identities also hold for M\o ller charges and currents, that is, if the replacements $\overline{\psi} \to \overline{\omega_-\phi_f}$ and $\psi \to \omega_+\phi_i$ are  made. In particular the M\o ller identities must hold at all orders in $e$, although verification order by order is tedious.  It follows immediately from the M\o ller analog of (\ref{VCUR}) that the concatenated vector M\o ller current $\int \overline{\omega_-\phi_f}\gamma^\mu \omega_+\phi_id\tau$ is divergenceless. 

The divergence of the axial M\o ller current  $\int \overline{\omega_-\phi_f}\gamma^\mu\gamma_5  \omega_+\phi_i d \tau$  is next recalculated correctly to ${\mathcal O}(e^2)$ for a particle in an external field. The particle is assumed for simplicity to be massless, that is, $m_e \to 0$\,. Fermion loops and triangles are again closed using the substitution (\ref{Gapprox}).  It is straightforward to show that the concatenated axial M\o ller divergence vanishes at ${\mathcal O}(1)$, and also at  ${\mathcal O}(e)$\,. At ${\mathcal O}(e^2)$ the divergence is the sum of three terms, each of which is an integral over the energy--momenta of photons at all the vertices of a triangle. The three terms are found to be identical, and so a symmetry factor of $S=3$ is assigned. The  assignment is not $S=3!$ since the factor of 2 is already incorporated into each of the three original integrals, following symmetrization with respect to the photons at the two vector vertices. Hence the required divergence is the value of just one such term, which is precisely equal to the divergence of the QED  coordinate--time--ordered, vacuum--to--vacuum amplitude of the product of an axial current and two vector currents \cite{Zee03}. The `anomalous' divergence is accordingly found to have the  standard value \cite{ Zee03, IZ,Sr07} 
\be\label{axanom}
\partial_\lambda J_5^\lambda(x)=-\frac{e^2}{(4\pi)^2}\varepsilon^{\mu\nu\rho\sigma}F_{\mu\nu}(x)F_{\rho\sigma}(x)\,.
\ee
Here, however, the fields are classically valued rather than operator valued.

\section{Summary and discussion}\label{S:summdisc}

\subsection{summary}\label{S:summ}
The parametrized Dirac equation  with a semiclassical Maxwell potential reproduces the standard one--loop corrections and axial anomaly of QED. The correct Feynman diagrams are the consequences here of the substitution (\ref{Gapprox}), which is justified for $\Delta p$ small. The correctness of the Feynman diagrams for arbitrary $\Delta p$ implies that the substitution is an effective partial summation of the perturbative series. Parametrized  relativistic quantum mechanics  provides, unlike field--theoretic QED,  an elementary and unrestricted representation of entanglement in space or time. The Bethe--Salpeter equation for bound states is a direct consequence of the parametrized Dirac equation for multiple particles.

\subsection{discussion}\label{S:disc}
The electromagnetic radiation considered here is semiclassical. That is, the electromagnetic four--potential  is a classical solution of Maxwell's equation possibly in the presence of a Dirac current. Semiclassical explanations of the photoelectric effect, variously attributed to Jaynes and to Lamb and Scully, are now accepted \cite{Fox06,GreZaj05} although informal controversy continues over the quantum mechanics of the photodetector. The semiclassical solutions of Maxwell's equation are typically subject to an homogeneous boundary condition, and so the classical vacuum is therefore a null potential.  The Rayleigh--Jeans law, with its ultraviolet catastrophe, is of course another prediction of the classical theory. The Quantum Theory of Radiation (QTR)  predicts \cite{IZ, Wei95, Zee03, Sr07} that the vacuum is a zero--point quantum field of energy density $\hbar\omega/2$ per mode of circular  frequency $\omega$, leading in particular to the Planck law or blackbody spectrum. Consider, however, Maxwell's equation subject to a boundary condition which is a Lorentz--invariant classical free potential,  which is statistically stationary and isotropic, and which has the same energy density  $\hbar\omega/2$ per mode. The classical solution can \cite{Boy69,Boy75,Camp99} then  account not only for the blackbody spectrum, but also for the Van der Waals forces, the Casimir effect and the Einstein `$A$' coefficient for spontaneous emission.  A random classical potential cannot \cite{Fox06} , however, explain antibunching for light beams passing through the two arms of an interferometer.  That is, owing to the Cauchy--Schwarz inequality, the correlation of intensity for classical beams must be greater than unity. Correlations less than unity are routinely observed \cite{Thorn04} at very low levels of illumination, and such antibunching can be explained \cite{Fox06} with QTR. It appears that only  QTR  can explain the sub-Poisson statistics which are detected again in very weak light beams \cite{Fox06}. Yet there is no immediate prospect of observing antibunching or sub--Poisson statistics for the $W$ or $Z$ gauge bosons, while a weak beam of free gluons is thus far a theoretical impossibility. 

It remains to consider the wider utility of parametrized relativistic quantum mechanics. The Higgs mechanism \cite{IZ}, for example, is essentially unaffected. The semiclassical massless gauge potential  is  independent of the parameter $\tau$, while the scalar Higgs \wf obeys the parametrized wave equation of Stueckelberg \cite{Stu41a,Stu41b,Stu42} with the standard `sombrero' self--interaction.

\section*{References}
\bibliography{paraDirac_Bib}

\end{document}